\documentclass[11pt]{article}
\usepackage{epsfig}
\usepackage{times}
\usepackage{cospar}
\usepackage[sectionbib]{natbib}
\title{PSR J2021+3651: A NEW $\gamma$-RAY PULSAR CANDIDATE}

\author{M.~S.~E. Roberts\address{Physics Department, McGill University, 
Rutherford Physics Building, 3600 University Street, Montreal, QC, H3A~2T8, 
Canada}$^{,}$\address{Center for Space Research, Massachusetts Institute of 
Technology, Cambridge, MA 02139, USA}, 
J.~W.~T. Hessels$^{1}$, 
S.~M. Ransom$^{1,2}$, 
V.~M. Kaspi$^{1}$, 
P.~C.~C. Freire\address{NAIC, Arecibo Observatory, HC03 Box 53995, PR 00612, USA}
, 
F. Crawford\address{Physics Department, Haverford College, Haverford, PA 19041,
 USA}, 
\\
and D.~R. Lorimer\address{University of Manchester, 
Jodrell Bank Observatory, Macclesfield, Cheshire SK11 9DL, UK}}

\begin{document}

\maketitle

\begin{abstract}
The {\it COS~B} high energy $\gamma$-ray source 
2CG~075+00, also known as GeV~J2020+3658 or 3EG~J2021+3716, has avoided
identification with a low energy counterpart for over twenty years.  
We present a likely identification with the discovery and subsequent 
timing of a young and energetic
pulsar, PSR J2021+3651, with the Wideband Arecibo Pulsar Processor at the 
Arecibo Observatory.  PSR~J2021+3651 has a rotation period $P \sim 104$~ms 
and $\dot P \sim 9.6\times10^{-14}$, implying a characteristic age 
$\tau_c \sim 17$ kyr and a spin-down luminosity 
$\dot E \sim 3.4\times 10^{36}$ erg~s$^{-1}$.  The pulsar is also coincident
with the {\it ASCA} source AX J2021.1+3651. The implied luminosity of the 
associated X-ray source suggests the X-ray emission is dominated by a 
pulsar wind nebula unresolved by {\it ASCA}.   
The pulsar's unexpectedly high dispersion measure 
(DM~$ \sim 371$~pc cm$^{-3}$) and the $d \ge 10$~kpc DM distance
pose a new question: is PSR J2021+3651 an extremely efficient $\gamma$-ray
pulsar at the edge of the Galaxy?  This is a question for {\it AGILE} and 
{\it GLAST} to answer.
\end{abstract}

\section*{INTRODUCTION}

Finding low energy counterparts to $\gamma$-ray sources is difficult because
they have large positional uncertainties, typically $\sim 1^{\circ}$
across.  Consequently, the majority of the high energy $\gamma$-ray sources 
observed by {\it COS B} and {\it EGRET} remain unidentified 
\citep{hbb+99}.  Young pulsars are viable candidates for many of these sources,
 and they remain the only discrete Galactic source class (other than the Sun) 
unambiguously shown to emit radiation in the 
100~MeV -- 10~GeV range. The increased sensitivity to pulsars
with high dispersion measure (DM), provided by new pulsar backends such as the
 Parkes multibeam system \citep{mlc+01}, has led to a number of discoveries 
recently of young pulsars coincident with known $\gamma$-ray
sources \citep{dkm+01_mal,cbm+01,hcg+01}. The recent detection of a young, 
low luminosity, radio pulsar
in the supernova remnant 3C58 \citep{csl+02} suggests many more faint 
radio pulsars 
await discovery in deep, targeted, searches. 

\begin{figure}
\begin{center}
\epsfig{file=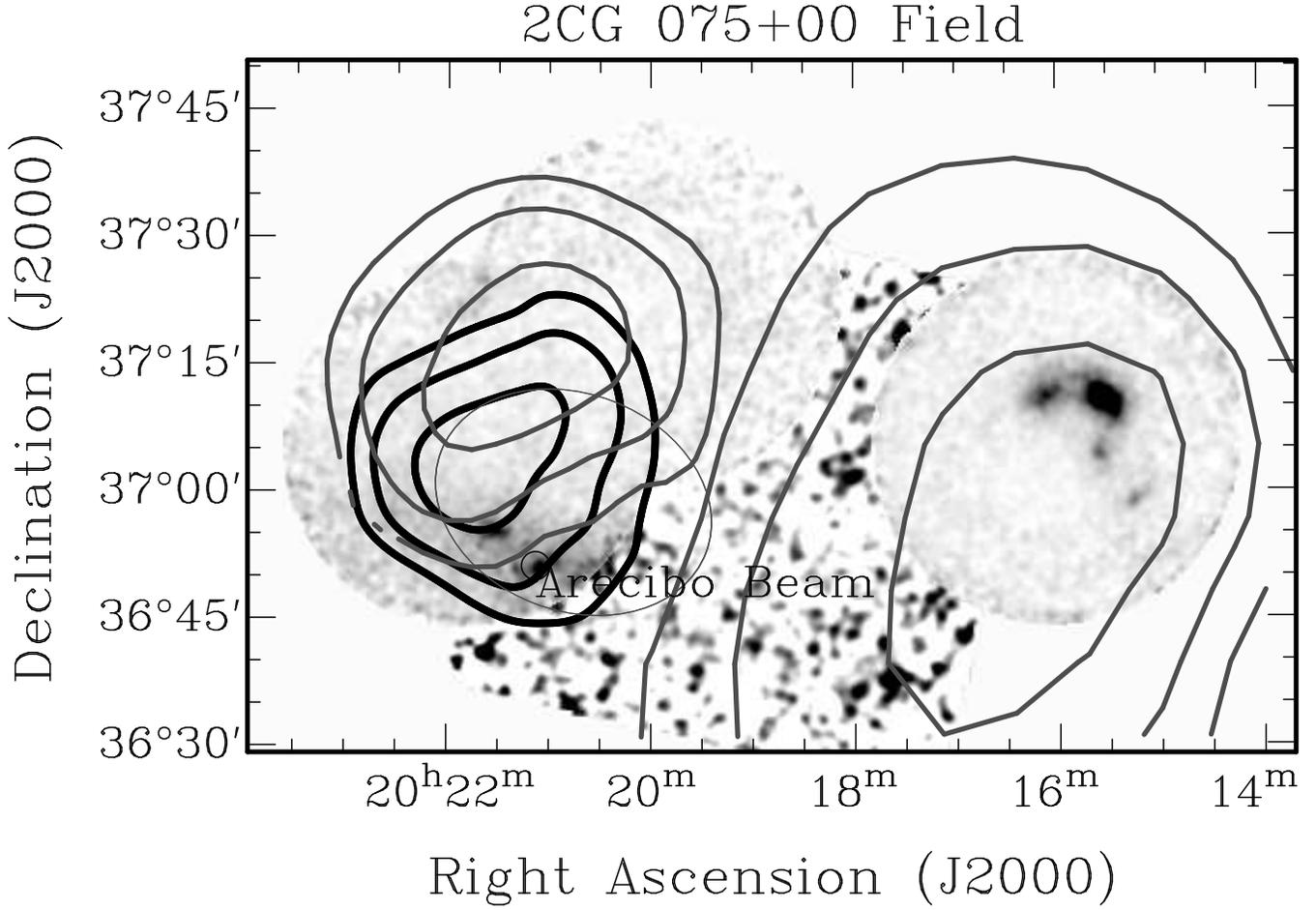,width=7in}
\end{center}
\caption{ \label{fig:xray} ~{\it ASCA} GIS 2--10~keV images of the {\it COS B} 
region 2CG 075+00 superimposed on an {\it Einstein} IPC 0.1--4.5~keV image.  
The light grey contours are the 68$\%$, 95$\%$, and 99$\%$ confidence regions 
of the third {\it EGRET} catalog sources 3EG~J2021+3716 (left) and 
3EG~J2016+3657 (right).  The dark contours are for the source GeV~J2020+3658 
and are derived from the $>$ 1~GeV photons only \citep{rrk01}.  The oval is 
the positional error derived for GeV~J2020+3658 by \citet{lm97}.  The circle 
centered on AX J2021.1+3651 indicates the size of the $3^{\prime}$ Arecibo 
beam.}
\end{figure}

The {\it COS B} source 2CG~075+00 has 
long been considered a pulsar candidate due to
its hard spectrum. {\it EGRET} resolved it into two sources.  The fainter one,
3EG J2016+3657, was identified as a probable blazar by \citet{muk+00}.
A {\it ROSAT} image and an {\it ASCA} GIS image based on the
second {\it EGRET} catalog position of the  brighter source, 3EG J2021+3716,
failed to yield any obvious candidates. These sources are in the 
crowded Cygnus region and the likelihood analysis used to find
$\gamma$-ray source positions is sensitive to the assumed number of 
surrounding sources. \citet{rrk01} (hereafter, RRK) noted that the 
catalog of sources above 1 GeV \citep{lm97} contained two sources in this
region that were not in the list of 3 $\sigma$ sources, based on 100 MeV
and above maps, used to construct the 3rd {\it EGRET} catalog. They 
rederived the positional contours of 3EG J2021+3716 using 1 GeV and above
maps and all of the known nearby sources in their analysis. An {\it ASCA}
image of the new position yielded two hard point-like sources (Figure~\ref{fig:xray}),
embedded in softer extended emission. One is identified with the 
Wolf-Rayet + O-star binary system WR 141. The second, AX J2021.1+3651,
is moderately absorbed (n$_{\rm H}$ = $5\times 10^{21} {\rm cm}^{-2}$) with
a hard, power-law spectrum (photon index $\Gamma=1.7$).  

By targetting the X-ray source AX J2021.1+3651 with the Arecibo radio 
telescope,
we have discovered the young and energetic 
pulsar PSR J2021+3651 \citep{rhr+02}. Here we present a timing solution 
of the pulsar based on eight months of monitoring with Arecibo, and 
discuss the likely prospect that this is the counterpart to 2CG~075+00, 
which has eluded identification for over twenty years.

\section*{OBSERVATIONS AND ANALYSIS}

PSR J2021+3651 was observed 17 times between MJD 52305 and 52545
using the Wideband Arecibo Pulsar Processor (WAPP) at the Arecibo Observatory.  
The WAPP is a fast-dump digital
correlator with adjustable bandwidth (50 or 100~MHz) and variable numbers 
of lags and sample times (for details see Dowd, Sisk, and Hagen 
2000\nocite{dsh00}).
Our observations were made at 1.4~GHz with 100~MHz of bandwidth, 512~lags, 
200~$\mu$s sampling, and summed 
polarizations.  The 16-bit samples were written to a disk array 
and then transfered to magnetic tape for later analysis.

\begin{figure}
\begin{center}
\epsfig{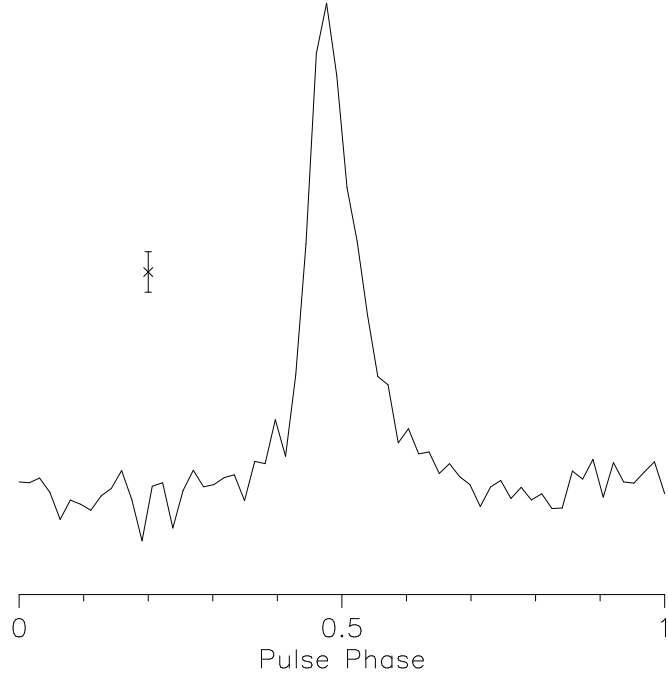}
\end{center}
\caption{\label{fig:pulse} ~1.4~GHz pulse profile for PSR J2021+3651 adopted 
from 
\citet{rhr+02}. The error bar represents the 1~$\sigma$ uncertainty.}
\end{figure}

Analysis of the observations was done using the {\tt PRESTO} software suite 
\citep{ran_thesis}\footnote{http://www.physics.mcgill.ca/$\sim$ransom/}.
Integrated pulse profiles (Figure~\ref{fig:pulse}) from our observations were convolved with a 
template profile to extract 69 topocentric times of arrival (TOA).  Using 
{\tt TEMPO}\footnote{See http://pulsar.princeton.edu/tempo}, the topocentric 
TOAs were converted to TOAs at the solar system barycenter at infinite 
frequency 
and fit simultaneously for pulsar period, period derivative, DM, RA, and Dec 
with a residual rms of 292~$\mu$s. 
Table~\ref{tab:pulsar} gives the measured and derived parameters for 
PSR J2021+3651, and is updated from \citet{rhr+02}.  The quoted errors are
three times those given by {\tt TEMPO} in order to compensate for systematic 
errors otherwise unacounted for.

\begin{table}
\begin{center}
\begin{minipage}{120mm}
\caption{~Measured and derived parameters for PSR~J2021+3651}
\label{tab:pulsar}
\begin{tabular}{l @{\hspace{2cm}} c}
\vspace{-4mm} \\
\hline
\vspace{-4mm} \\
Parameter & Value \\
\vspace{-4mm} \\
\hline
\vspace{-4mm} \\
Right ascension $\alpha$ (J2000) & 20$^{\rm{h}}$ 21$^{\rm{m}}$05$^{\rm{s}}$.214(51)\\
Declination $\delta$ (J2000) & +36$^\circ$ 51$'$ 08$''$.42(72) \\
Galactic longitude $l$          & 75.$^{\circ}$23 \\
Galactic latitude $b$          & +0.$^{\circ}$11 \\
Pulse period $P$ (s)         & 0.103722225646(12) \\
Period derivative  $\dot P$     & $ 9.56118(60) \times 10^{-14}$  \\  
Pulse frequency $\nu$ (s$^{-1}$) & 9.6411351933(11) \\         
Frequency derivative  $\dot{\nu}$ (s$^{-2}$) & $-8.88726(54) \times 10^{-12}$ \\
RMS residual & 292$\mu$s \\
Epoch (MJD)        & 52407.389 \\
Dispersion Measure DM (pc cm$^{-3}$) & 369.12(58) \\
Pulse width at 50$\%$ of peak w$_{50}$ (ms)       & 9.9  \\
Pulse width at 10$\%$ of peak w$_{10}$ (ms)       & 18 \\
Flux density at 1425 MHz (mJy)     & $\sim 0.1$ \\                             
Spin-down luminosity $\dot{E}$$^a$ (erg s$^{-1}$) & $3.4 \times 10^{36}$ \\
Surface dipole magnetic field $B$$^b$ (G)  & $3.2 \times 10^{12}$ \\
Characteristic Age $\tau_{c} \equiv \frac{1}{2} P / \dot P$ (kyr)     &  17 \\
\vspace{-4mm} \\
\hline
\vspace{-4mm} \\
\end{tabular}
$^a$ $\dot E = 4 {\pi}^2 I \dot P / P^3$ with $I = 10^{45}$ g cm$^2$ \\
$^b$ Assuming standard magnetic dipole spindown:  
$B = 3.2 \times 10^{19} (P \dot P)^{1/2}$ \\ Gauss \citep{mt77}

\end{minipage}
\end{center}
\end{table}

\section{DISCUSSION}

The DM of PSR J2021+3651 is by far the highest known in the 
Galactic longitude 
range $55^{\circ} < l < 80^{\circ}$, which is mainly an inter-spiral arm 
direction. Using the recent \citet{cl02} update to the \citet{tc93} 
dispersion measure model gives a distance
of $\sim 12.4$ kpc, at the outer edge of the last spiral arm used in the 
model. 
It is possible that there are further contributions
from clouds in the Cygnus region not included in the model, 
where there is known to be excess gas at $d\sim 1.5$~kpc (J. Cordes,
private communication, 2002).  However, there are no obvious HII regions 
within the Arecibo beam seen in either Very Large Array (VLA) 20-cm radio or
Midcourse Space Experiment (MSX) 8.3~$\mu$m images (available from 
the NASA/IPAC Infrared Science Archive).  

The likelihood that PSR J2021+3651
is unrelated to the X-ray source is extremely small given the rarity 
of such young, 
energetic, pulsars and the consistency of the timing position with the 
$\sim 1'$ positional uncertainty of the {\it ASCA} source.
However, the high DM is somewhat surprising given the X-ray absorption quoted by RRK (n$_{\rm H}$=(5.0$\pm0.25)\times 10^{21}$~cm$^{-2}$), where the errors
represent the 90\% confidence region.
The total Galactic HI column density in this direction as estimated from the
FTOOL {\tt nh}, which uses the HI map of \citet{dl90}, is $1.2\times 10^{22}$~cm$^{-2}$.  This should be
a good approximation if the source is truly at the far edge of the 
outer spiral arm. Noting that the {\it ASCA} image shows faint, softer emission
in the region (Figure~\ref{fig:xray}), and given the likely possibility of either 
associated thermal X-ray flux from a supernova remnant or a nearby massive star, we fit the {\it ASCA} spectrum of RRK,
adding a thermal component to the absorbed power-law model.
Accounting for $\sim4$\% of the photon flux with a MEKAL thermal plasma model 
(see \citet{log95} and references therein)
of temperature $kT\sim 0.1$~keV in {\tt XSPEC} \citep{arn96} statistically
improves the fit ({\it F}-test chance probability of 2.5\%). The
best-fit absorption for this three component model is 
n$_{\rm H}$=7.6$\times 10^{21}$~cm$^{-2}$ with a 90\% confidence region
of (4.1 -- 12.3)$\times 10^{21}$~cm$^{-2}$, consistent
with the total Galactic column density. The best-fit photon index
is $\Gamma=1.86$, still consistent with the 1.47 -- 2.01 range in RRK
derived from the simple absorbed power-law model.  Hence the X-ray
absorption does not force us to adopt a smaller distance than is suggested
by the DM and the value of n$_{\rm H}$ may be contaminated by thermal emission.
Upcoming {\it Chandra} observations of the source may well clarify this.

For a distance $d_{10}=d/10$~kpc, the inferred
isotropic X-ray luminosity $L_{\rm X}=4.8\times 10^{34} d_{10}^2$ erg s$^{-1}$ (2 -- 10~keV). 
The X-ray efficiency $\eta_{\rm X}=L_{\rm X}/\dot E$ is 0.01$d^2_{10}$.
Compared to 
the total pulsar plus nebula X-ray luminosity of other spin-powered pulsars,
this is somewhat high, but within the observed scatter
\citep{pccm02,che00}. Upcoming Chandra observations will determine what
fraction of the X-ray flux comes from a compact nebula, and will allow
us to search for X-ray pulsations.

2CG~075+00 = 3EG J2021+3716 = GeV J2020+3658 is the 10th brightest 
Galactic source above 1 GeV. It has a hard (photon index $\Gamma=1.86\pm
0.10$) spectrum with no sign of a break out to 10 GeV, and has low
variability, similar to the known $\gamma$-ray pulsars. 
PSR J2021+3651's position is outside the 99\% contour of the 3EG catalog
position, but consistent with the 95\% contour of both the \citet{lm97}
GeV catalog position and the RRK position. The only other potential 
counterpart within any of the error contours is WR 141. Although such binary
systems have the potential for producing $\gamma$-rays from colliding
winds, no convincing association with a known high-energy ($>$ 100~MeV) 
$\gamma$-ray source has
yet been made for a Wolf-Rayet system \citep{rbt99}. 
The pulsar's positional coincidence coupled with the high inferred spin-down 
luminosity strongly suggests this pulsar emits $\gamma$-rays.
Unfortunately, confirming this by folding archival 
{\it EGRET} data is problematic due to
the likelihood of significant past timing noise and glitches, 
which make the back-extrapolation of the rotational ephemeris uncertain. 

Assuming the distances given by the \citet{cl02} model, the known
$\gamma$-ray pulsars have $\gamma$-ray efficiencies 
$\eta_\gamma=L_{\gamma}/\dot E$ mostly between 0.0001 and 0.03 
(assuming 1 sr beaming).  
Assuming the $\gamma$-rays are 100\% pulsed, the 
inferred $\gamma$-ray efficiency for PSR J2021+3651,
$\eta_\gamma =0.15 d_{10}^2$ in the 100~MeV to 10~GeV range, would 
be by far the most efficient $\gamma$-ray pulsar.  Note however that PSR B1055$-$52 \citep{tbb+99} has a similar efficiency if one uses the DM distance given by \citet{tc93}, instead of the newer \cite{cl02} model.
PSR B1706$-$44, which has nearly identical spin parameters, has a 
$\gamma$-ray efficiency of $\sim 0.01$, while Vela with a
similar spin-down energy has a low efficiency of $\sim 0.0004$.
Even if the pulsar is located within the Perseus arm at 
a distance of 5~kpc, it would still be very efficient. If it is
the $\gamma$-ray source, it  would
further strain the relationship of efficiency to spin-down energy
$\eta_{\gamma}\propto L_{sd}^{-1/2}$ expected from theory \citep{zh00}.
While there is currently no observational evidence for 
a closer distance, increased DM from an
intervening source in this relatively crowded direction would not
be surprising. We note that the DM derived distances for several other
young pulsars
recently discovered within {\it EGRET} error boxes, if the
true counterparts, also tend to have high inferred $\gamma$-ray efficiencies \citet{rob02}. 

We plan to continue timing observations of this source, which will refine the position, provide a
contemporaneous timing solution for future high-energy pulse 
searches, and allow us to monitor for glitches.  {\it AGILE} and {\it GLAST} should be able to unambiguously determine                   
if the pulsar is the counterpart of the $\gamma$-ray source.

\section*{ACKNOWLEDGEMENTS}

We thank Jim Cordes for useful discussions.  We acknowledge support from 
NSERC, CFI, an NSF CAREER Award, and a Sloan Fellowship.  
M.S.E.R. is a Quebec Merit fellow.  S.M.R. is a Tomlinson fellow.  
J.W.T.H. is an NSERC PGS~A fellow. V.M.K. is a Canada Research Chair.  
The Arecibo Observatory is part of the National Astronomy and Ionosphere 
Center, which is operated by Cornell University under a cooperative 
agreement with the National Science Foundation.

\vspace{0.5cm}
\noindent
E-mail address of J.W.T. Hessels: hessels@physics.mcgill.ca \\
Manuscript received 19 October 2002


\begin{thebibliography}{100}
\parskip=0pt                                                                   
\parsep=0pt 
\itemsep=0pt


\bibitem[{Arnaud(1996)}]{arn96}
Arnaud, K.~A., XSPEC: The first ten years, in {\it Astronomical Data Analysis Software and Systems V}, eds.
  G.~Jacoby and J.~Barnes, {\it A.S.P. Conference Series}, {\bf 101}, 17-20, 1996.

\bibitem[Camilo {et~al.}(2001)]{cbm+01}
Camilo, F., Bell, J.~F., Manchester, R.~N. et al., PSR J1016-5857: A young radio pulsar with possible supernova remnant, X-ray, and Gamma-ray associations, {\it Astrophys.~J.~L.}, {\bf 557}, L51-L55, 2001.

\bibitem[{{Camilo} {et~al.}(2002){Camilo}, {Stairs}, {Lorimer}, {Backer},
  {Ransom}, {Klein}, {Wielebinski}, {Kramer}, {McLaughlin}, {Arzoumanian}, and
  {M{\" u}ller}}]{csl+02}
Camilo, F., Stairs, I.~H., Lorimer, D.~R. et al., Discovery of radio pulsations from the X-ray pulsar J0205+6449 in supernova remnant 3C 58 with the Green Bank Telescope, {\it Astrophys.~J.~L.}, {\bf 571}, L41-L44, 2002.

\bibitem[{{Chevalier}(2000)}]{che00}
{Chevalier}, R.~A., A model for the X-ray luminosity of pulsar nebulae, {\it Astrophys.~J.}, {\bf 539}, L45-L48, 2000.

\bibitem[Cordes (2002)]{} Cordes,~J.~M.  Private communication, 2002.

\bibitem[Cordes and Lazio(2002)]{cl02} Cordes, J.~M. and Lazio, T.~J.~W., NE2001.I. A new model for the Galactic distribution of free electrons and its fluctuations, {\it astro-ph/0207156}, 2002.

\bibitem[{{D'Amico} {et~al.}(2001){D'Amico}, {Kaspi}, {Manchester}, {Camilo},
  {Lyne}, {Possenti}, {Stairs}, {Kramer}, {Crawford}, {Bell}, {McKay},
  {Gaensler}, and {Roberts}}]{dkm+01_mal}
D'Amico, N., Kaspi, V.~M., Manchester, R.~N. et al., Two young radio pulsars coincident with EGRET sources, {\it Astrophys.~J.~L.}, {\bf 552}, L45-L48, 2001.

\bibitem[{Dickey and Lockman(1990)}]{dl90}
Dickey, J.~M. and Lockman, F.~J., HI in the Galaxy, {\it Ann.~Rev.~Astron.~Astrophys.}, {\bf 28}, 215-261, 1990.

\bibitem[{{Dowd} {et~al.}(2000){Dowd}, {Sisk}, and {Hagen}}]{dsh00}
Dowd, A., Sisk, W., and Hagen, J., WAPP -- Wideband Arecibo Pulsar Processor, in {\it ASP Conf. Ser. 202: IAU Colloq. 177: Pulsar Astronomy - 2000 and Beyond}, 275, 2000.

\bibitem[{{Halpern} {et~al.}(2001){Halpern}, {Camilo}, {Gotthelf}, {Helfand},
  {Kramer}, {Lyne}, {Leighly}, and {Eracleous}}]{hcg+01}
Halpern, J.~P., Camilo, F., Gotthelf, E.~V. et al., PSR J2229+6114: Discovery of an energetic young pulsar in the error box of the EGRET source 3EG J2227+6122, {\it Astrophys.~J.~L.}, {\bf 552}, L125-L128, 2001.

\bibitem[Hartman et al.(1999)]{hbb+99} Hartman, R.~C., Bertsch, D.~L., Bloom, S.~D. et al., The third EGRET catalog of high-energy Gamma-ray sources, {\it Astrophys.~J.~Suppl.}, {\bf 123}, 79-202, 1999.
 
\bibitem[Lamb and Macomb(1997)]{lm97} Lamb, R.~C.~and Macomb, 
D.~J., Point sources of GeV gamma rays, {\it Astrophys.~J.}, {\bf 488}, 872-880, 1997.

\bibitem[Liedahl, Osterheld, and Goldstein(1995)]{log95}           
Liedahl, D.~A., Osterheld, A.~L., and Goldstein, W.~H., New calculations of Fe L-shell X-ray spectra in high-temperature plasmas, {\it Astrophys.~J.}, {\bf 438}, L115-L118, 1995.
    
\bibitem[{Manchester {et~al.}(2001)Manchester, Lyne, Camilo, Bell, Kaspi,
  D'Amico, McKay, Crawford, Stairs, Possenti, Morris, and Sheppard}]{mlc+01}
Manchester, R.~N., Lyne, A. G., Camilo, F. et al., The Parkes multi-beam pulsar survey - I. observing and data analysis systems, discovery and timing of 100 pulsars, {\it Mon.~Not.~R.~Astron.~Soc.}, {\bf 328}, 17-35, 2001.

\bibitem[{Manchester and Taylor(1977)}]{mt77}
Manchester, R.~N. and Taylor, J.~H., Pulsars (San Francisco: Freeman), 1977.

\bibitem[{{Possenti} {et~al.}(2002){Possenti}, {Cerutti}, {Colpi}, and
  {Mereghetti}}]{pccm02}
Possenti, A., Cerutti, R., Colpi, M. et al., Re-examining the X-ray versus spin-down luminosity correlation of rotation powered pulsars, {\it Astron.~Astroph.}, {\bf 387}, 993-1002, 2002.

\bibitem[{{Mukherjee} {et.~al.}(2000)}]{muk+00}
Mukherjee, R., Gotthelf, E.~V., Halpern, J. et al., Multiwavelength examination of the COS B field 2CG~075+00 yields a blazar identification for 3EG J2016+3657, {\it Astrophys.~J.}, {\bf 542}, 740-749, 2000.

\bibitem[{{Ransom}(2001)}]{ran_thesis}
Ransom, S.~M., New search techniques for binary pulsars, Ph.D.~Thesis, Harvard, 2001.

\bibitem[Roberts, Romani, and Kawai(2001)]{rrk01}
Roberts, M. S.~E., Romani, R.~W., and Kawai, N., The ASCA catalog of potential X-ray counterparts of GEV sources, {\it Astrophys.~J.~Suppl.}, {\bf 133}, 451-465, 2001.

\bibitem[{{Roberts} {et~al.}(2002a) {Roberts}, {Hessels}, {Ransom}, {Kaspi}, {Freire}, {Crawford}, and {Lorimer}}]{rhr+02}
Roberts, M.~S.~E., Hessels, J.~W.~T., Ransom, S.~M. et al., PSR J2021+3651: A young radio pulsar coincident with an unidentified EGRET $\gamma$-ray source, {\it Astrophys.~J.~L.}, {\bf 577}, L19-L22, 2002.

\bibitem[{{Roberts} (2002b)}]{rob02}
Roberts, M.~S.~E., Pulsar searches of unidentified EGRET sources, {\it astro-ph/0212080}, 2002 .

\bibitem[Romero, Benaglia, and Torres(1999)]{rbt99}
Romero, G.~E., Benaglia, P., and Torres, D.~F., Unidentified 3EG gamma-ray sources at low galactic latitudes, {\it Astron. and Astrophys.}, {\bf 348}, 868-876, 1999.

\bibitem[{Taylor and Cordes(1993)}]{tc93}
Taylor, J.~H. and Cordes, J.~M., Pulsar distances and the Galactic distribution of free electrons, {\it Astrophys.~J.}, {\bf 411}, 674-684, 1993.

\bibitem[{Thompson {et~al.}(1999)Thompson, Bailes, Bertsch, Cordes, D'Amico,
  Esposito, Finley, Hartman, Hermsen, Kanbach, Kaspi, Kniffen, Kuiper, Lin,
  Manchester, Matz, Mayer-Hasselwander, Michelson, Nolan, Ogelman, Pohl,
  Ramanamurthy, Sreekumar, Reimer, Taylor, and Ulmer}]{tbb+99}
Thompson, D.~J., Bailes, M., Bertsch, D. L. et al., Gamma radiation from PSR B1055-52, {\it Astrophys.~J.}, {\bf 516}, 297-306, 1999.

\bibitem[{Zhang and Harding(2000)}]{zh00}
Zhang, B. and Harding, A.~K., Full polar cap cascade scenario: Gamma-ray and X-ray luminosities from spin-powered pulsars, {\it Astrophys.~J.}, {\bf 532}, 1150-1171, 2000.

\end{thebibliography}
\end{document}